# Title: Signatures of Fractional Quantum Anomalous Hall States in Twisted MoTe₂ Bilayer

**Authors:** Jiaqi Cai[1†], Eric Anderson[1†], Chong Wang[2], Xiaowei Zhang[2], Xiaoyu Liu[2], William Holtzmann[1], Yinong Zhang[1], Fengren Fan[3,4], Takashi Taniguchi[5], Kenji Watanabe[6], Ying Ran[7], Ting Cao[2], Liang Fu[8], Di Xiao[2,1], Wang Yao[3,4], Xiaodong Xu[1,2*]

[1]Department of Physics, University of Washington, Seattle, Washington 98195, USA
[2]Department of Materials Science and Engineering, University of Washington, Seattle, Washington 98195, USA
[3]Department of Physics, University of Hong Kong, Hong Kong, China
[4]HKU-UCAS Joint Institute of Theoretical and Computational Physics at Hong Kong, China
[5]International Center for Materials Nanoarchitectonics, National Institute for Materials Science, 1-1 Namiki, Tsukuba 305-0044, Japan
[6]Research Center for Functional Materials, National Institute for Materials Science, 1-1 Namiki, Tsukuba 305-0044, Japan
[7]Department of Physics, Boston College, Chestnut Hill, MA 02467, USA
[8]Department of Physics, Massachusetts Institute of Technology, Cambridge, Massachusetts 02139, USA
† These authors contributed equally to the work.
*Corresponding author's email: xuxd@uw.edu

**Abstract:** The interplay between spontaneous symmetry breaking and topology can result in exotic quantum states of matter. A celebrated example is the quantum anomalous Hall (QAH) state, which exhibits an integer quantum Hall effect at zero magnetic field thanks to its intrinsic ferromagnetism[1-3]. In the presence of strong electron-electron interactions, exotic fractional-QAH (FQAH) states at zero magnetic field can emerge. These states could host fractional excitations, including non-Abelian anyons – crucial building blocks for topological quantum computation[4,5]. Flat Chern bands are widely considered as a desirable venue to realize the FQAH state[6-9]. For this purpose, twisted transition metal dichalcogenide homobilayers in rhombohedral stacking have recently been predicted to be a promising material platform[10-13]. Here, we report experimental signatures of FQAH states in ~3.7-degree twisted MoTe₂ bilayer. Magnetic circular dichroism measurements reveal robust ferromagnetic states at fractionally hole filled moiré minibands. Using trion photoluminescence as a sensor[14], we obtain a Landau fan diagram which shows linear shifts in carrier densities corresponding to the $v = -2/3$ and $-3/5$ ferromagnetic states with applied magnetic field. These shifts match the Streda formula dispersion of FQAH states with fractionally quantized Hall conductance of $\sigma_{xy} = -\frac{2}{3}\frac{e^2}{h}$ and $-\frac{3}{5}\frac{e^2}{h}$, respectively. Moreover, the $v = -1$ state exhibits a dispersion corresponding to Chern number -1, consistent with the predicted QAH state[10-13]. In comparison, several non-ferromagnetic states on the electron doping side do not disperse, i.e., are trivial correlated insulators. The observed topological states can be further electrically driven into topologically trivial states. Our findings provide clear evidence of the long-sought FQAH states, putting forward MoTe₂ moiré superlattices as a fascinating platform for exploring fractional excitations.

**Main Text:**

The fractional quantum anomalous Hall (FQAH) state is the zero magnetic field analog of the fractional quantum Hall state. While both are strongly correlated topological phases of matter[6-9,15-18], the FQAH state additionally requires spontaneous time-reversal symmetry breaking without Landau level formation[6,7,9]. Despite extensive theoretical studies, experimental realization of the FQAH state remains challenging due to the lack of a suitable physical system. A promising approach is to realize the Haldane model on a honeycomb lattice with strong correlation, leading to topological flat bands that share key characteristics of Landau levels[19,20]. Moiré superlattices of two-dimensional materials are emerging platforms for this approach, thanks to their highly tunable flat electronic bands[12], topology[11], lattice geometry[13,21], and correlation effects[22]. Experimentally, the integer quantum anomalous Hall (QAH) state, a precursor to the FQAH, has been realized in both graphene[23-25] and twisted transition metal dichalcogenides (TMD) systems[26]. Furthermore, fractional Chern insulator states have also been observed in Bernal staked bilayer graphene/hBN superlattices[27] with a magnetic field near 30T and in magic-angle twisted bilayer graphene[28] with magnetic fields as low as 5T. All progress points to the 2D moiré superlattice as a powerful platform for engineering flat Chern bands with spontaneous time reversal symmetry breaking at fractional fillings.

Recent theoretical works[10-13,29,30] have suggested that rhombohedral-stacked TMD moiré bilayer can host topological flat bands with opposite Chern numbers in the two spin/valley sectors. Both integer and fractional QAH states have been predicted in these systems. Indeed, in near 4-degree twisted $MoTe_2$ homobilayers, electrically tunable ferromagnetic states that spontaneously break time-reversal symmetry have been observed in a wide range of hole doping phase space[21], with a Curie temperature as high as 14 K at integer filling $v = -1$ (one hole per moiré unit cell). Building on this progress, in this work, we provide evidence of FQAH states at zero magnetic field in a fractionally filled moiré superlattice in twisted $MoTe_2$ bilayer.

We fabricated high-quality R-stacked $MoTe_2$ bilayer samples with a twist angle of ~3.7 degrees. Figure 1a illustrates the standard dual gate device geometry, which enables independent control of carrier density $n$ and perpendicular electric field $D$ in the sample. The R-stacked bilayer hosts two degenerate energy minima forming a honeycomb lattice (Fig. 1b). The complex hopping between the sublattice sites within the same layer (i.e., next nearest neighbor hopping in the honeycomb lattice) can open a topological band gap when the honeycomb lattice is filled with two holes per cell ($v = -2$), realizing a Kane-Mele model[11,13]. At filling $v = -1$ of the honeycomb lattice, ferromagnetism arises due to strong exchange interactions[21] and lifts the spin-valley flavor degeneracy of the effective Kane-Mele model, realizing a Haldane model with strong Coulomb interactions, i.e., a Haldane-Hubbard model[12,19].

Below, we first establish robust ferromagnetism corresponding to correlated insulating states at fractional fillings. We then present measurement of topological invariants associated with the $v = -2/3$ and $-3/5$ fractionally filled states using an optically detected Landau fan diagram, where evidence of FQAH states is obtained by comparison to the Streda formula dispersion. Lastly, we demonstrate that these FQAH states can be switched on and off by exploiting electric field control of lattice geometry, and thus the magnetic ground states and band topology.

## Ferromagnetic states at fractionally filled moiré minibands

Figure 1c shows the photoluminescence (PL) intensity plot versus $n$ and photon energy (see Extended Data Fig. 1 for the full doping range). All data are taken at a temperature of ~1.6 K, unless otherwise specified. For doping-dependent measurements, we focus on the hole doping side since the electron side is found to be non-ferromagnetic and topologically trivial. At $v = -1$ and $-2/3$, PL intensity drops while the peak blue shifts. A weak feature at $v = -3/5$ is also observed under a magnetic field of ~1T (Extended Data Fig. 1). As reported[21], this PL resonance arises from the trion (i.e. charged exciton). The observed reduction in PL intensity is a result of reduced trion population, as the formation of correlated insulating states depletes the holes available for trion formation.

To investigate the magnetic response of the twisted bilayer, we employ reflective magnetic circular dichroism (RMCD) measurements. Figure 1d is the zero magnetic field RMCD signal intensity plot as a function of $v$ and $D$ (see Extended Data Fig. 2 for the full doping range and additional spots). The red area of non-vanishing RMCD signal highlights the phase space of the ferromagnetic states, consistent with the previous report[21]. Interestingly, the data show that the critical electric field $D_c$ for suppressing the ferromagnetic state is enhanced at the fractionally filled insulating state $v = -2/3$. Figure 1e displays the hysteresis of the RMCD signal versus magnetic field $\mu_o H$ at selected $v$. Sharp spin-flip transitions at critical field (i.e., coercive field) $H_C$ are observed.

The dependence of hysteresis loops and $H_C$ on the correlated insulating states is further highlighted by measurements of the RMCD hysteresis loop as a function of doping. Figure 2a shows the RMCD intensity plots versus $v$ and $\mu_o H$ swept down (left) and up (right). The difference between the sweeps is indicated by the intensity plot of the hysteretic component (Fig. 2b). The data clearly show that the coercive field $\mu_o H_C$ is enhanced near the correlated insulating states. As hole doping increases from charge neutrality, $\mu_o H_C$ first grows from ~10 mT to ~20 mT near the $v = -2/3$ insulating state, then reduces slightly, and subsequently increases drastically up to ~100 mT near the $v = -1$ insulting state before finally reducing until the ferromagnetic state vanishes.

Figure 2c is a high-resolution plot of the RMCD hysteresis around the fractional filling features. The extracted coercive field $H_C$ and Curie temperature $T_C$ versus $v$ are shown in Fig. 2d. In addition to the robust $v = -2/3$ state, a weak and spatially dependent feature exists near $v = -3/4$ filling in $H_C$ (Extended Data Fig.3). As an additional probe of the strength of the ferromagnetic states, we perform temperature-dependent RMCD measurements versus filling (Extended Data Fig. 4). The extracted Curie temperature also has a strong dependence on filling factors, peaking at $v = -2/3$ and $-1$ where correlated insulating states form, with a $T_C$ of ~4.5K and ~14K, respectively. This $T_C$ is relatively high considering the large moiré unit cell of period ~5.4 nm and correspondingly low spin density compared to conventional 2D magnets.

## Measurement of topological invariants

Topological invariants associated with gapped incompressible states can be measured via Hall conductivity or with a fan diagram of charge gaps fitted to the Streda formula. In the latter approach, the topological invariant is $C = \phi_o \frac{\partial n}{\partial B}$, where $\phi_o$ is the magnetic flux quantum, $n$ is the carrier density of the gapped state, $B$ is the magnetic field, and $C$ is equal to the Hall conductance (in units of $e^2/h$). The Streda formula approach[28,31] is particularly suitable when electrical transport measurement is inconvenient. In the fan diagram, if the state is topologically trivial, the carrier density should be independent of the magnetic field. However, the carrier density of gapped topologically nontrivial states should shift linearly versus magnetic field with a quantized slope corresponding to the Chern number.

We have employed a trion sensing technique to optically measure the fan diagram. A similar trion sensing technique has been used to image fractional quantum Hall liquids in GaAs/AlGaAs quantum wells[14]. Excitons in TMDs are very sensitive to changes in the local dielectric environment[32,33]. Thus, they have been used to probe charge gaps of a variety of correlated states, such as generalized Wigner crystal states in TMD moiré superlattices[34-37], as well as fractional quantum Hall states in graphene[38]. As shown in Fig. 1c, trion PL intensity drops and its peak energy blue shifts when correlated insulating states form. From this signature, we can determine the carrier density corresponding to the $v = -1$ ($n_{-1}$) and $-2/3$ ($n_{-2/3}$) correlated insulating states. We then measure the density $n_{-1}$ and $n_{-2/3}$ as a function of magnetic field, probing the topological nature of these ferromagnetic insulating states.

Figure 3a shows PL intensity plots versus carrier density at selected $\mu_o H$, for $D = 0$. The data clearly show that $n_{-1}$ and $n_{-2/3}$ shift linearly as $\mu_o H$ increases, with $n_{-1}$ changing with a larger slope than that of $n_{-2/3}$. We extract the spectrally integrated PL intensity and plot it as a function of magnetic field and carrier density in Fig. 3b (i.e., the optically detected Landau fan diagram, see Methods). We observe that the $v = -1$ and $-2/3$ states show linear dispersions that persist down to zero magnetic field. In addition, there is a weak dispersive feature at $v = -3/5$. This feature becomes appreciable as $\mu_o H$ increases. These observations clearly differ from the topologically trivial correlated insulating states at the electron doping side (Fig. 3d and Extended Data Fig. 5), which are nondispersive. This result aligns with our expectations, as these states at the electron doping side are not ferromagnetic - a necessary condition for the formation of both integer and fractional QAH states. Figure 3c is the Wannier diagram with solid lines indicating dispersion curves determined by the Streda formula. We find that Streda dispersions with $C = -1$ (black line), $-2/3$, and $-3/5$ (blue lines) match well with the $n$-$\mu_o H$ dependence of three correlated ferromagnetic insulating states in Fig. 3b at $v = -1$, $-2/3$, and $-3/5$, respectively.

The observed $C = -1$ state supports the assignment of the $v = -1$ ferromagnetic insulating state as a QAH state (or a topological Mott insulator[19]). This observation is consistent with recent predictions from the Haldane Hubbard model[1,10-12,19,39] with a Chern number of $-1$. Most significantly, the ferromagnetic insulating states with $C = -2/3$ and $-3/5$ at fractional fillings of the flat Chern band are identified as FQAH states, i.e., fractional Chern insulator states which survive to zero magnetic field. These two states are the zero-field analog of the fractional quantum Hall state sequence of odd denominators, hosting fractional charge excitations[40] and Abelian topological orders. We have also performed exact diagonalization calculations based on realistic material parameters, the results of which are consistent with the two observed FQAH states. Figure 3e shows that the three-fold degenerate ground states at $v = -2/3$ evolve into each other upon flux insertion (see Supplementary Text for details of the theoretical model and calculations for $v = -3/5$ state). The many-body Chern numbers for $v = -2/3$ and $-3/5$ are also found to be $C = -2/3$ and $-3/5$, respectively.

**Electrically controlled topological phase diagram**

Because the sublattice orbitals of the honeycomb lattice are localized in opposite layers of the twisted MoTe$_2$ bilayer, the application of a perpendicular electric field breaks the layer degeneracy. This enables tuning of the superlattice from a honeycomb to a triangular lattice dominated by kinetic antiferromagnetic exchange interactions. Our previous study demonstrated a ferromagnetic to antiferromagnetic phase transition concurrent with such a change in lattice geometry[21]. Thus, we anticipate this moiré geometry control will fundamentally change the topological properties of the moiré bands and therefore the many-body ground states at integer and fractional fillings.

To illustrate this electric field control of topology, we set the $D$ field to -250 mV/nm. We then measure the PL spectra versus $n$ at selected magnetic fields. As shown in Fig. 4a, only the correlated insulating state at $v = -1$, a Mott state in the triangular lattice, survives at large $D$. There is no signature of the $v = -2/3$ state for this large $D$ field. The carrier density corresponding to the $v = -1$ state becomes independent of magnetic field, demonstrating that the state is now a nondispersive and topologically trivial insulator. Figure 4b shows the RMCD signal versus $D$ at zero magnetic field. As expected, RMCD signal and thus ferromagnetism vanish at large $D$, consistent with the observed topologically trivial state.

Because the magnetic behavior of the correlated states underpins their topological properties, we examine the magnetic phase diagram of the system as a function of filling, electric field, and temperature. Figure 4c shows the RMCD signal intensity plot versus $v$ and $D$ at a temperature of 3.5K (see Extended Data Fig. 6 for hysteresis loops at selected fillings). At this temperature, the ferromagnetic states become less robust to the variation of $v$ and $D$ compared to the 1.6K case. We found the ferromagnetic phase spaces near $v = -1$ and $v = -2/3$ to be disconnected. Though the magnetic states away from $v = -2/3$ soften significantly compared to their behavior at 1.6K, the $v = -2/3$ state remains a hard magnet with $\mu_o H_C \sim 15$mT. This observation further supports the view that the correlated insulating state at $v = -2/3$ thermodynamically stabilizes ferromagnetism. Figure 4d shows the RMCD signal intensity plot of the $v = -2/3$ state as a function of temperature and $D$ at zero magnetic field, i.e., the magnetic phase diagram of the $v = -2/3$ state. The $D$ field range of the FM state decreases as temperature increases, eventually vanishing above a $T_C$ of about 4.5K. Both temperature and electric field can destroy the ferromagnetic and thus FQAH states, highlighting the high tunability of the twisted MoTe$_2$ bilayer system.

**Conclusions**

Our work provides experimental signatures of two FQAH states in twisted MoTe$_2$ bilayer in R-stacking. The discovered -2/3 and -3/5 FQAH states are expected to host Abelian anyon excitations with fractional statistics. With improved sample quality, it is possible that non-Abelian anyons - the key component for realizing topological quantum computation - may be observed in other fractionally filled states[4]. There have been many theoretical developments regarding FQAH states over the past several decades[6-10,41-45]. Our work provides an experimental playground to test some of these ideas with unprecedented tunability via electric field, doping, temperature, and magnetic field. An immediate next step is to probe the topological properties using electrical transport measurements. As the spin-valley degree of freedom can be accessed using circularly polarized light, it is also possible to realize optical control of magnetization[46] and thus FQAH (and QAH) states. More importantly, our work indicates a feasible path to exploit the remarkable properties and tunability of TMD moiré superlattices for quantum engineering of exotic topological orders arising from spontaneous symmetry breaking and strong correlation effects.

## Methods

### Device fabrication

Devices were fabricated using the tear-and-stack method. First, hBN, used as the gate dielectric, and graphite, used for the metallic gates, were mechanically exfoliated onto Si/SiO$_2$ substrates. Homogenous flakes were identified using an optical microscope. hBN thickness was confirmed by atomic force microscope measurements. In an argon filled glovebox with H$_2$O and O$_2$ concentrations < 0.1ppm, 2H MoTe$_2$ (HQ Graphene) was mechanically exfoliated onto Si/SiO$_2$ substrates with a 285nm oxide layer precleaned by oxygen plasma. Monolayer flakes were identified via optical microscope. Standard polymer-based dry transfer techniques were used to fabricate the heterostructure. First, the topmost hBN protection layer, top graphite electrode, hBN top dielectric layer and half of the MoTe$_2$ monolayer were picked up. Then, the entire stage was rotated by the desired twist angle. After rotation, the second half of the MoTe$_2$ monolayer was picked up, forming the twisted MoTe$_2$ interface. The heterostructure was completed by picking up a long strip of graphite to serve as a grounding pin, a bottom hBN dielectric, and the graphite bottom gate. The transfer process was performed at ~100 °C. The entire heterostructure was then put down and picked up repeatedly at ~140 °C to mechanically squeeze out trapped gas bubbles between layers, before being melted down onto the substrate at ~170°C. The polymer was dissolved in anhydrous chloroform for 5 minutes in an inert glovebox environment. Standard electron beam lithography was used to electrically connect gold wire bonding pads to the gates and grounding pin of the fully encapsulated device. Liftoff was performed in an inert glovebox environment with anhydrous dichloromethane. Compared to the earlier report of the observation of ferromagnetic states in this system[21], we found that a large homogeneous area with reduced moiré disorder is crucial for the observations reported in this work (see Extended Data Fig. 2).

### Optical measurements

All measurements were performed in a closed-loop magneto-optical exchange gas cryostat (attoDRY 2100) with an attocube *xyz* piezo stage, 9T out-of-plane superconducting magnet, and base temperature of 1.6K. For photoluminescence, a 632.8 nm HeNe laser was focused on the sample by a high-NA nonmagnetic objective to minimize the magnetic field induced drift. Differential reflectance measurements were performed under the same conditions using a Tungsten-Halogen lamp. The reflected signal was dispersed with a diffraction grating (Princeton Instruments, 600 grooves/mm at 1 μm blaze) and detected by a LN cooled infrared CCD (Princeton Instruments PyLoN-IR 1.7). For photoluminescence, a long pass filter was inserted before the entrance to the spectrometer to remove reflected laser excitation.

RMCD was performed by filtering a broadband supercontinuum source (NKT SuperK EXTREME EXW-12) by dual-passing through a monochromator, achieving a narrow excitation bandwidth in resonance with the trion feature. The out-of-plane magnetization of the sample induces an MCD signal ΔR, the difference between the reflected right- and left-circularly polarized light. To obtain the normalized RMCD ΔR/R, the laser intensity was modulated by a chopping frequency of $p$ = 850Hz and the phase was modulated by λ/4 via a photoelastic modulator at a frequency of $f$ = 50kHz. The output signal of an InGaAs avalanche photodiode detector was read by two lock-in amplifiers (SR830). The ratio between the *p*-component signal $I_p$ and *f*-component signal $I_f$ gives the RMCD signal: $\Delta R/R = I_f /(J_1(\pi/2) \times I_p)$ where $J_1$ is the first-order Bessel function. All zero field RMCD plots are taken after the device was trained by application of a positive magnetic field, unless otherwise specified.

**Determination of doping density and electric field**

The carrier density $n$ and electric field $D$ on the sample are converted from top (bottom) gate voltage $V_{tg}$ ($V_{bg}$) using a parallel plate capacitor model: $n = (V_{tg}C_{tg} + V_{bg}C_{bg})/e - n_{offset}$ and $D/\varepsilon_0 = (V_{tg}C_{tg} - V_{bg}C_{bg})/2\varepsilon_0 - D_{offset}$, where $C_{tg}$ and $C_{bg}$ are the top and bottom gate capacitance obtained from the device geometry, $e$ is the electron charge, and $\varepsilon_0$ is the vacuum permittivity. The offset carrier density is derived from fitting to the integer and fractional states in PL spectra. The offset electric field is determined from the symmetric axis of the dual gate RMCD map. The obtained doping density from the capacitor model can then be used to calculate twist angle from the assigned filling factors in the optical measurements, which is comparable to the targeted twist angle.

**Determination of Chern number**

The optical Landau fan diagram (Fig. 3b) was obtained by integrating PL spectra over a 1 meV spectral range around the peak. To determine the Chern number, first, using the doping density given by hBN capacitance ($C_g = \varepsilon_{hBN}/d$, $\varepsilon_{hBN} = 3.0$) and assuming the filling to be linear in carrier density, the Landau fan diagram from the PL measurement can be reproduced. The hBN thickness is measured using an atomic force microscope (AFM) with an uncertainty of ~200 pm, which is negligible compared to the 20-40 nm hBN. The most significant uncertainty comes from the dielectric constant of hBN (widely accepted as 3-3.3). Second, single-gated reflectance measurements are performed to obtain the capacitance ratio of the top and bottom gates, which matches the thickness ratio obtained by AFM measurements. Using the slope extracted from the Landau fan diagram, the capacitance of the gates can be obtained and compared to the hBN capacitance. This method cross-checks the validity of the hBN capacitance value.

**Acknowledgements:** We thank Xi Wang for helpful discussion. CW acknowledges helpful discussions with Yuchi He and Heqiu Li. This project is mainly supported by the U.S. Department of Energy (DOE), Office of Science, Basic Energy Sciences (BES), under the award DE-SC0018171. RMCD measurements are partially supported by Air Force Office of Scientific Research (AFOSR) Multidisciplinary University Research Initiative (MURI) program, grant no. FA9550- 19-1-0390. Measurements of electrical control of topological phase transitions are partially supported by AFOSR FA9550-21-1-0177. The device fabrication is partially supported by the Center on Programmable Quantum Materials, an Energy Frontier Research Center funded by DOE BES under award DE-SC0019443. The authors also acknowledge the use of the facilities and instrumentation supported by NSF MRSEC DMR-1719797. EA and WH acknowledge the support by the National Science Foundation Graduate Research Fellowship Program under Grant No. DGE-2140004. FF and WY acknowledge support from the Research Grants Council of Hong Kong SAR (AoE/P-701/20, HKU SRFS2122-7S05) and Croucher Foundation. CW and DX acknowledge the support from DoE BES under the award DE-SC0012509. K.W. and T.T. acknowledge support from the JSPS KAKENHI (Grant Numbers 19H05790, 20H00354 and 21H05233). XX acknowledges support from the State of Washington funded Clean Energy Institute and from the Boeing Distinguished Professorship in Physics.

**Author contributions:** XX conceived and supervised the experiment. JC fabricated the devices. TT and KW provided hBN crystals. EA and JC performed the magneto-optical measurements with help from WH and YZ. JC, EA, and XX analyzed the data and interpreted the results. LF contributed to the idea of trion sensing. CW, XZ, XL, FF, LF, YR, TC, DX, and WY provided theoretical support. JC, EA, LF, DX, WY, and XX wrote the manuscript with input from all authors. All authors discussed the results.

**Competing interests:** XX, EA and JC have applied for a patent partially based on this work. The other authors declare no competing interests.

**Data availability:** Source data that reproduces the plots are provided with this paper. All supporting data for this paper and other findings of this study are available from the corresponding author upon reasonable request.

Figures:

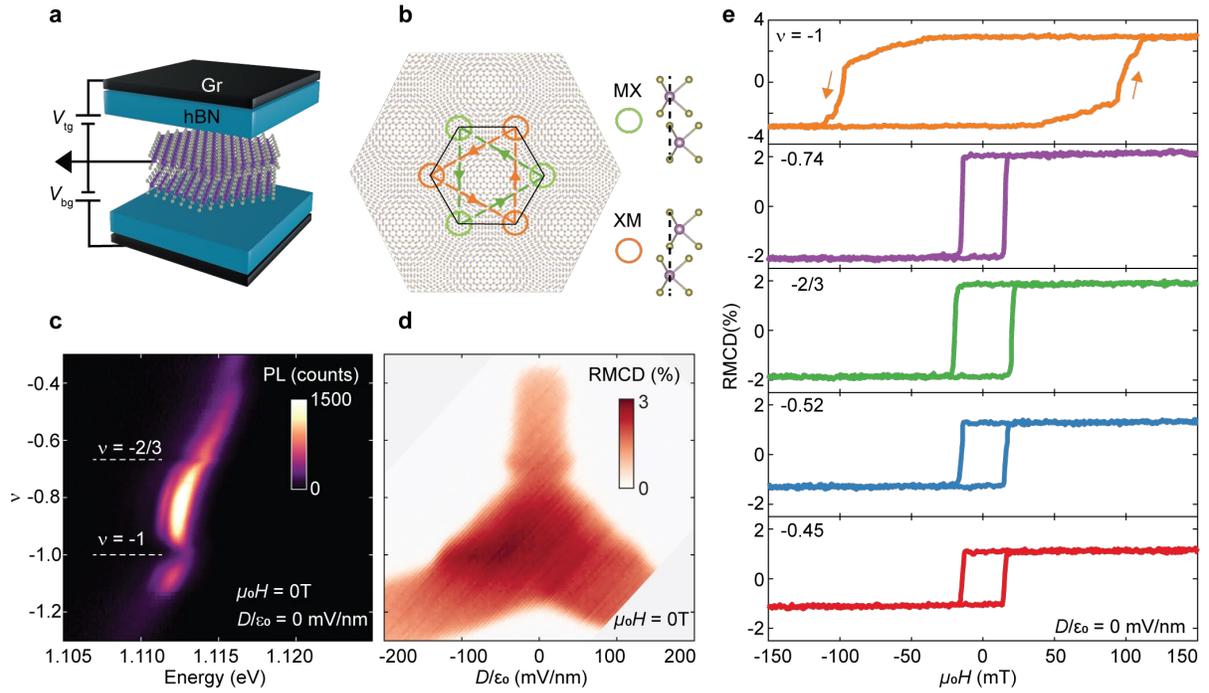

**Figure 1 | Electrically tunable correlated ferromagnetic states in twisted bilayer MoTe$_2$.**
**a,** Schematic of dual gated device structure. **b,** R-stacked homobilayer hosting two degenerate energy minima at high symmetry MX and XM points, forming a honeycomb moiré superlattice. Complex hopping between next-nearest-neighbor sites realizes the Haldane model. **c,** Photoluminescence intensity plot as a function of hole doping and photon energy. Filling factors $\nu$ corresponding to the formation of correlated insulating states are indicated. **d,** Reflective magnetic circular dichroism (RMCD) signal versus $\nu$ and perpendicular electric field $D$ at zero magnetic field $\mu_0 H$. The phase space with non-vanishing signal corresponds to the ferromagnetic state. **e,** RMCD signal versus $\mu_0 H$ swept back and forth at selected fillings.

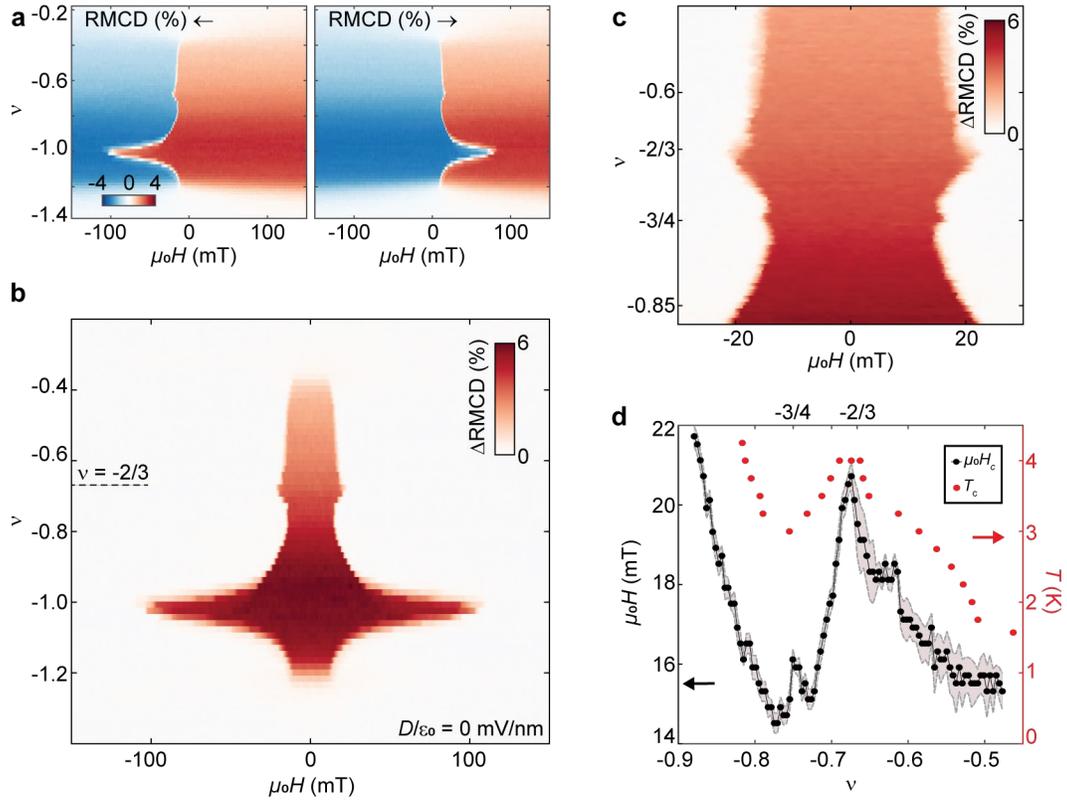

**Figure 2 | Fractionally filled correlated ferromagnetic insulating states. a,** RMCD signal intensity plot versus filling factor $v$ and magnetic field swept down (left) and up (right). **b,** Difference of RMCD sweeps in **a**, ΔRMCD, versus $v$ and $\mu_o H$. Clear enhancement of $H_C$ at $v$ = -1 and -2/3 is observed. **c,** High resolution plot of ΔRMCD, versus $v$ and $\mu_o H$. In addition to the -2/3 feature, another feature near -3/4 filling is observed. **d,** $H_C$ (left axis) and $T_C$ (right axis) versus $v$. Enhancement of both $H_C$ and $T_C$ is apparent near -2/3 filling. The value $H_C$ was extracted by change point detection. The shaded area denotes the extracted transition width. The extracted $T_C$ has an uncertainty of 0.25K, determined by the temperature step size of the measurement.

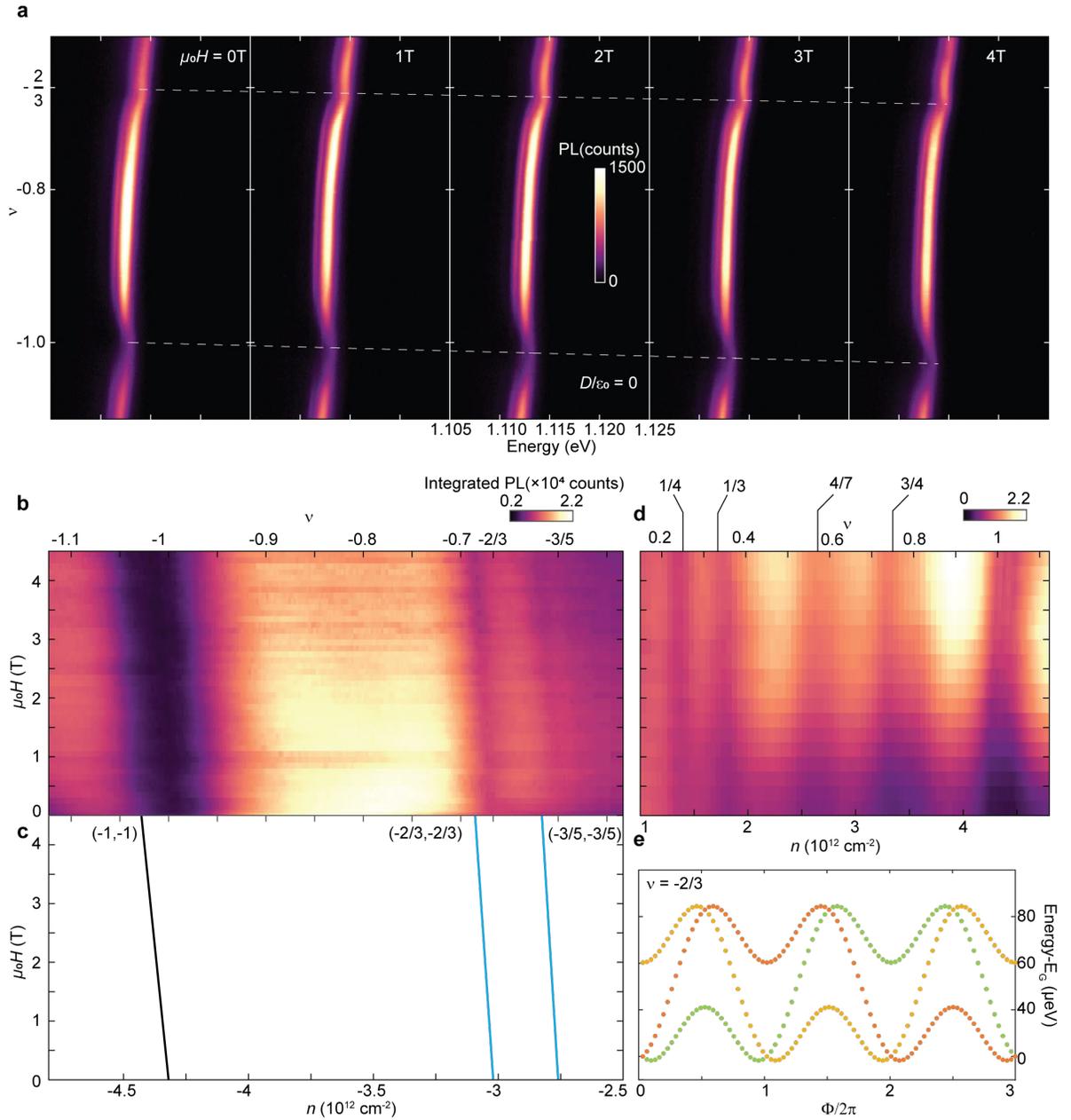

**Figure 3 | Evidence of integer and fractional quantum anomalous Hall states. a,** PL intensity plot versus $v$ and photon energy at selected magnetic fields. Dashed lines are guides to the eye. **b,** Spectrally integrated PL intensity versus $\mu_o H$ and carrier density $n$, i.e. an optical Landau fan diagram. **c,** Wannier diagram corresponding to a $C = -1$ QAH state at $v = -1$ (black line), $C = -2/3$ fractional QAH (FQAH) state at $v = -2/3$, and $C = -3/5$ FQAH state at $v = -3/5$ (blue lines), with $C$ equal to the Hall conductance in the units of $e^2/h$. States are marked by $(C,v)$. **d,** Integrated PL intensity versus $\mu_o H$ and $n$ showing correlated insulating states at electron filling. These states are dispersionless with magnetic field and topologically trivial. **e,** Exact diagonalization calculation of the ground states at $v = -2/3$, showing evolution of the three nearly degenerate ground states upon flux insertion.

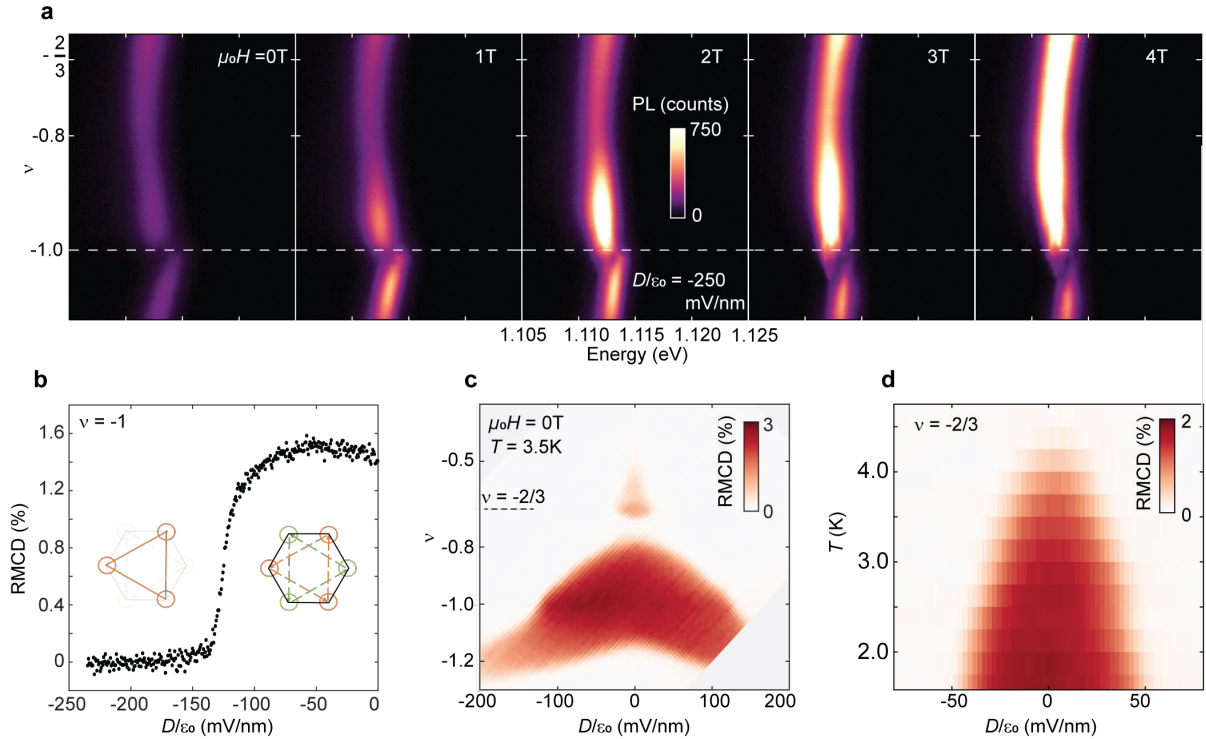

**Figure 4 | Electrically tunable topological phase transition. a,** As in Fig.3a, but with a large applied electric field of $D/\varepsilon_0$ = -250 mV/nm. The $\nu$ = -1 state becomes dispersionless and thus topologically trivial. **b**, Zero-magnetic field RMCD signal at $\nu$ = -1 versus electric field, demonstrating an electric field driven magnetic phase transition concurrent with the change in superlattice geometry from honeycomb to triangular (insets). **c,** RMCD signal intensity plot versus $\nu$ and $D$ at $T$ = 3.5K. **d,** RMCD signal versus temperature and $D$ at $\nu$ = -2/3, highlighting the ferromagnetic phase diagram.

# Extended Data Figures for

# Signatures of Fractional Quantum Anomalous Hall States in Twisted MoTe$_2$ Bilayer

**Authors:** Jiaqi Cai[1†], Eric Anderson[1†], Chong Wang[2], Xiaowei Zhang[2], Xiaoyu Liu[2], William Holtzmann[1], Yinong Zhang[1], Fengren Fan[3,4], Takashi Taniguchi[3], Kenji Watanabe[4], Ying Ran[5], Ting Cao[2], Liang Fu[6], Di Xiao[2,1], Wang Yao[7,8], Xiaodong Xu[1,2*]

[1]Department of Physics, University of Washington, Seattle, Washington 98195, USA
[2]Department of Materials Science and Engineering, University of Washington, Seattle, Washington 98195, USA
[3]Department of Physics, University of Hong Kong, Hong Kong, China
[4]HKU-UCAS Joint Institute of Theoretical and Computational Physics at Hong Kong, China
[5]International Center for Materials Nanoarchitectonics, National Institute for Materials Science, 1-1 Namiki, Tsukuba 305-0044, Japan
[6]Research Center for Functional Materials, National Institute for Materials Science, 1-1 Namiki, Tsukuba 305-0044, Japan
[7]Department of Physics, Boston College, Chestnut Hill, MA 02467, USA
[8]Department of Physics, Massachusetts Institute of Technology, Cambridge, Massachusetts 02139, USA
† These authors contributed equally to the work.
*Corresponding author's email: xuxd@uw.edu

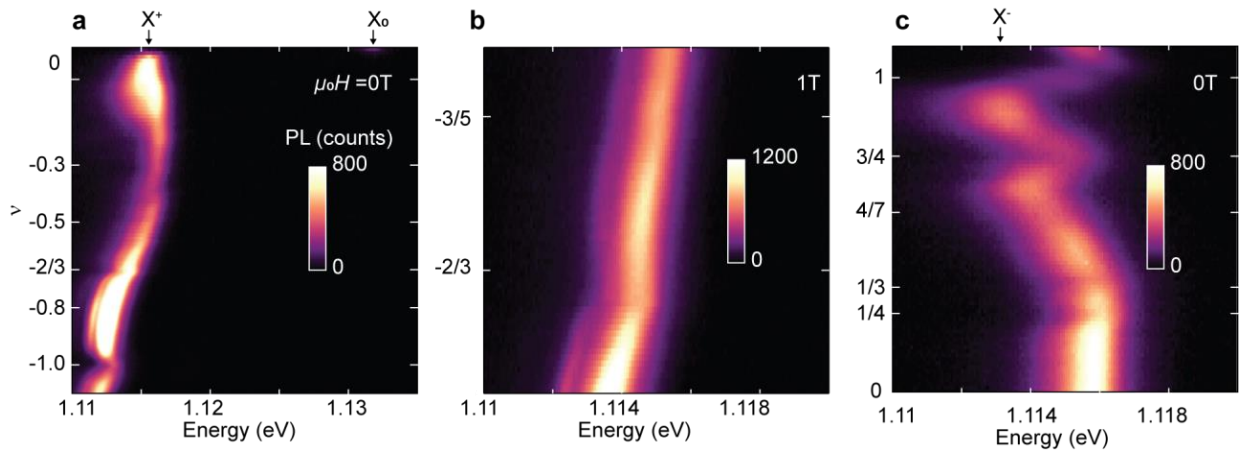

**Extended Data Figure 1 | Doping dependent photoluminescence. a,** Photoluminescence (PL) intensity plot as a function of photon energy and filling factor for hole doping. $X_0$: neutral exciton; $X^+$: positively charged trion. Two significant trion energy shifts with reduced PL intensity are visible at $v = -2/3$ and $v = -1$. **b,** Zoomed-in plot of PL at $\mu_o H = 1$T near fractional fillings for hole doping. An additional state at $v = -3/5$ appears as a trion energy shift with reduced PL intensity. **c,** PL intensity plot for electron doping. Multiple states including $v = 1, 3/4, 4/7, 1/3,$ and $1/4$ can be observed. $X^-$ denotes negatively charged trion.

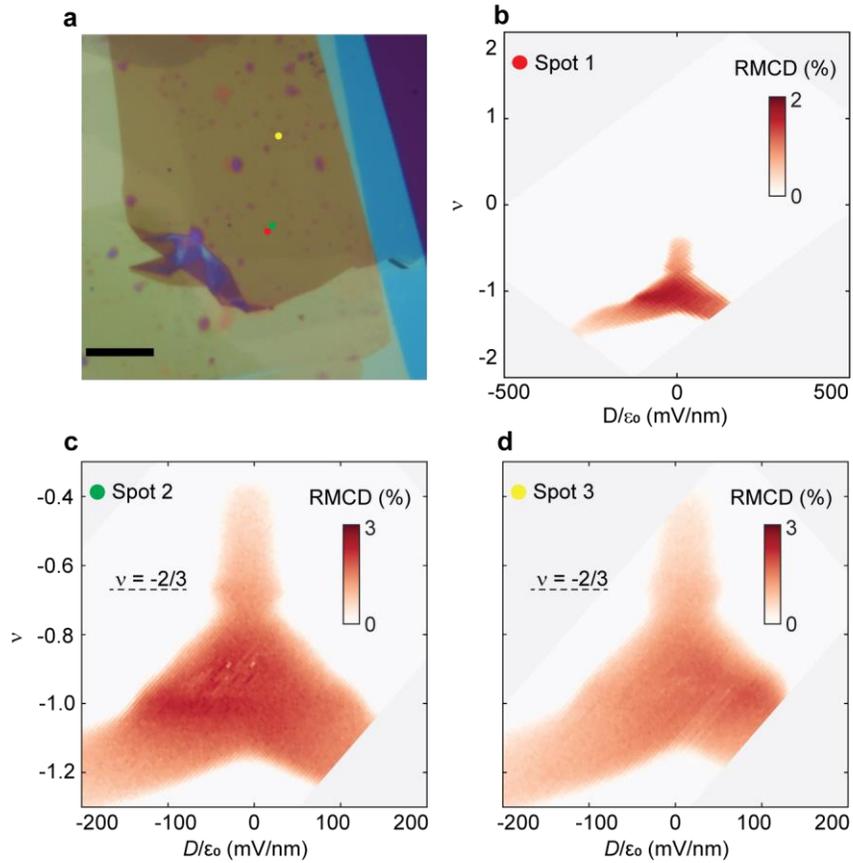

**Extended Data Figure 2 | Device image and position independence of correlated states. a,** Optical microscope image of the device. Scale bar: 10 µm. RMCD and PL data in the main text are taken in the homogenous region at spots 1 and 2, respectively. **b,** Zero magnetic field RMCD vs filling factor and displacement field at spot 1. The electron-doping side is non-ferromagnetic. **c-d,** additional RMCD vs $\nu$ and $D$ plots in spots 2 and 3. Both show an enhancement of the ferromagnetic state at $\nu = -2/3$, demonstrating repeatability.

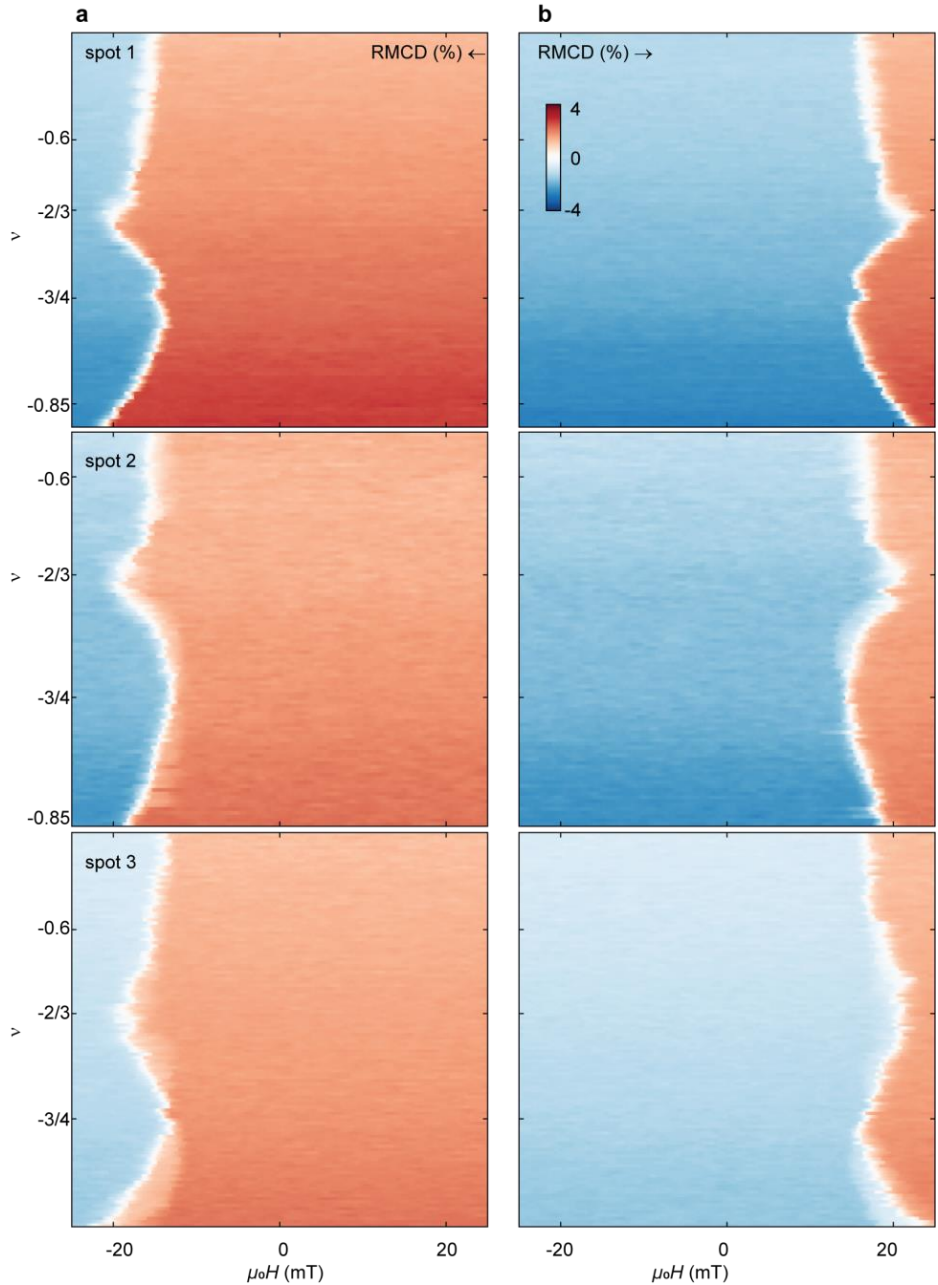

**Extended Data Figure 3 | Doping dependent RMCD hysteresis. a-b,** RMCD intensity plot versus filling factor $v$ and magnetic field $\mu_0 H$ swept down (**a**) and up (**b**) for spots 1, 2, and 3 as defined in Extended Data Fig. 2. For all spots, $\mu_0 H_C$ is enhanced near $v = -2/3$. However, the $v = -3/4$ state is not visible for spot 2, showing its spatial dependence.

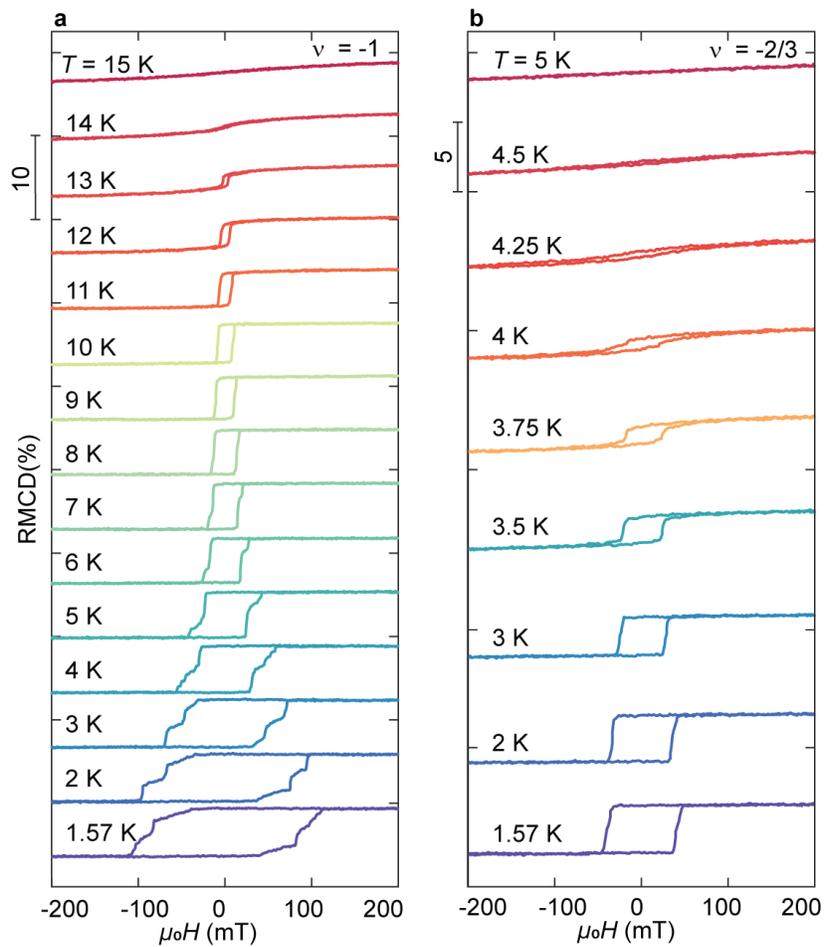

**Extended Data Figure 4 | Temperature-dependent RMCD at $v$ = -1 and -2/3. a,** $v$ = -1. The Curie temperature $T_C$ is ~14K. **b,** $v$ = -2/3. $T_C$ is ~4.5K.

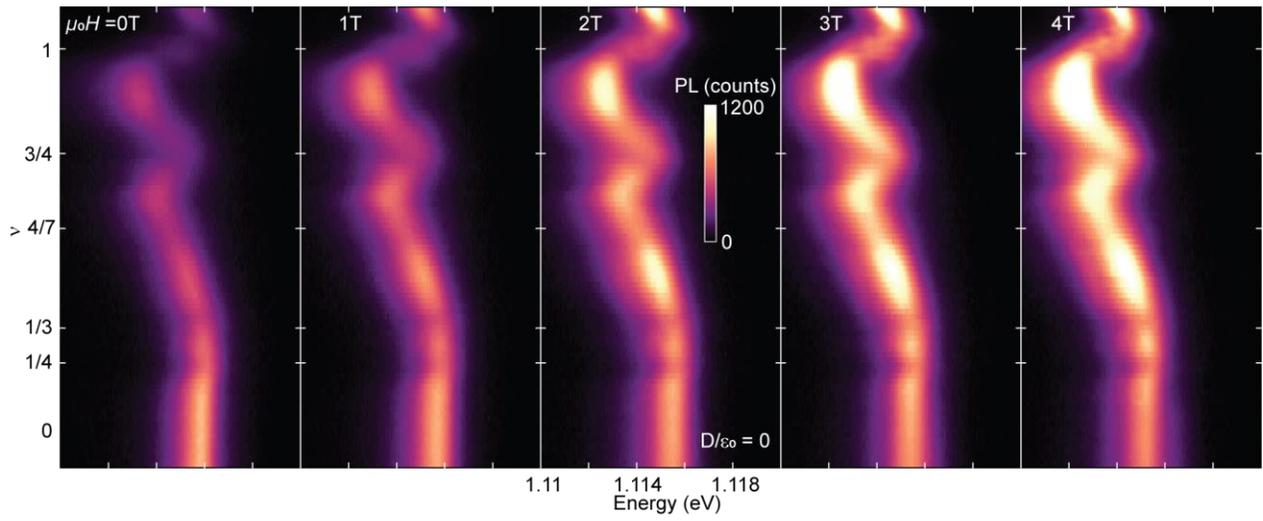

**Extended Data Figure 5 | Electron doping dependent photoluminescence.** PL vs electron doping at selected magnetic fields. No dispersion is observed for any correlated insulating states, implying that the lowest moiré conduction band is topologically trivial.

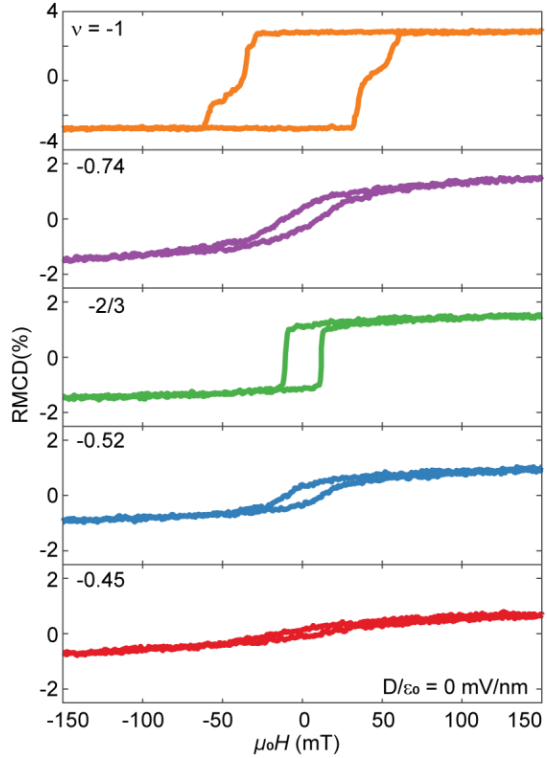

**Extended Data Figure 6 | RMCD hysteresis sweeps for selected fillings at $T = 3.5K$.** The ferromagnetic states at $v = -2/3$ and $v = -1$ retain substantial remnant RMCD signal and sharp spin-flip transitions, signatures of a hard magnet, at an elevated temperature. States at intermediate dopings, however, show behavior consistent with a soft magnet.